\documentclass[twocolumn,aps,prl,amssymb,amstext,showpacs]{revtex4}

\usepackage{graphicx,amsfonts,amssymb,amsmath,bm}
\usepackage{color}

\newcommand{\vx}{{\bf x}}
\newcommand{\vm}{{\bf m}}
\newcommand{\vH}{{\bf H}}

\newcommand{\vcH}{\bm{\mathcal{H}}}

\newcommand{\ud}{\mathrm{d}}

\newcommand{\ex}{{\mathrm{\bf e}_{x}}}
\newcommand{\erho}{{\mathrm{\bf e}_{\rho}}}
\newcommand{\epsi}{{\mathrm{\bf e}_{\psi}}}

\begin{document}

\title{Domain wall motion in ferromagnetic nanotubes: Analytic
  results}

\author{Arseni Goussev$^{1,2}$, JM Robbins$^3$, Valeriy Slastikov$^3$}

\affiliation{$^1$Department of Mathematics and Information Sciences,
  Northumbria University, Newcastle Upon Tyne, NE1 8ST, United
  Kingdom\\ $^2$Max Planck Institute for the Physics of Complex
  Systems, N{\"o}thnitzer Stra{\ss}e 38, D-01187 Dresden,
  Germany\\ $^3$School of Mathematics, University of Bristol,
  University Walk, Bristol, BS8 1TW, United Kingdom}

\date{\today}

\begin{abstract}
  Dynamics of magnetization domain walls (DWs) in thin ferromagnetic
  nanotubes subject to longitudinal external fields is addressed
  analytically in the regimes of strong and weak
  penalization. Explicit functional forms of the DW profiles and
  formulas for the DW propagation velocity are derived in both
  regimes. In particular, the DW speed is shown to depend nonlinearly 
 on the nanotube radius.
\end{abstract}

\pacs{75.75.-c, 75.78.Fg}

\maketitle

The problem of controlled manipulation of magnetization domains in
quasi-one-dimensional ferromagnetic nanostructures is of paramount
technological importance in designing new generation memory devices
\cite{Parkin08,Hayashi08,Thomas10} and of fundamental interest in the
vibrant areas of micromagnetics and spintronics. To date, substantial
theoretical progress has been achieved in understanding the dynamics
of domain walls (DWs) in nanowires under the influence of applied
magnetic fields
\cite{SchryerWalker74,KH04Magnetization,MCA+07Domain,Bryan08,Yang08, Wang09a, Wang09b, Lu10, SS10Fast,GRS10Domain,GLRSS13a, GLRSS13b}
and spin-polarized electric currents
\cite{Be78, Be84, Be96, Slon96, TNM+05Micromagnetic,YKG+10Beating,TA10Current,TLA12Domain, GLRSS13b}. Nevertheless,
the search for schemes and regimes allowing fast and energy efficient
DW propagation actively continues.

Recently, ferromagnetic nanotubes have been proposed as a guide of DWs
driven by an external magnetic field \cite{YAK+11Fast}. The central
advantage of this approach is the absence of the so-called Walker
breakdown that is unavoidable at sufficiently strong applied fields
for wire and strip geometries \cite{SchryerWalker74}. The absence of
the Walker breakdown is a topological effect which leads to a
significant increase of the DW stability and propagation speed
\cite{YAK+11Fast, GGRS11}.

In this paper, we analytically address the DW dynamics in thin
ferromagnetic nanotubes under the action of an external magnetic field
and derive an explicit formula for the DW propagation speed in the
regimes of strong and weak penalization. Our formula reveals a
nonlinear dependence of the propagation speed on the nanotube radius,
and may be used as a guide in devising new 
experiments.

\begin{figure}[ht]
\includegraphics[width=3in]{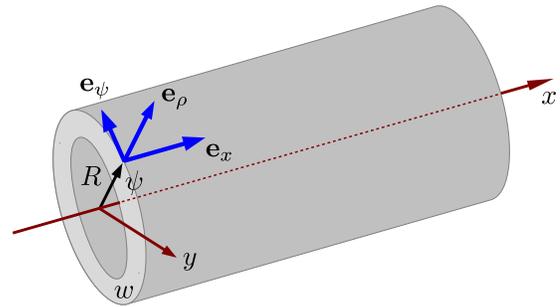}
\caption{(Color online) A sketch of a nanotube with an outer radius
  $R$ and an inner radius $R-w$. A point on the outer surface of the
  nanotube is parametrized by the coordinate $x$ along its symmetry
  axis and the polar angle $\psi$. The unit vectors $\ex$ (parallel to
  the symmetry axis), $\epsi$ (tangential to the surface), and $\erho$
  (normal to the surface) form a right-handed triplet.}
\label{fig1}
\end{figure}

We consider an infinitely long ferromagnetic nanotube with an outer
radius $R$ and an inner radius $(R-w)$ (see Fig.~\ref{fig1}). The
magnetization distribution at a spatial point $\vx$ and time $t$ is
described by ${\bf M}(\vx,t) = M_s \vm(\vx,t)$, where $|\vm(\vx,t)| =
1$ if $\vx \in \Omega$ (the point belongs to the nanotube region) and
$|\vm(\vx,t)| = 0$ if $\vx \not\in \Omega$ (the point lies outside the
nanotube region). Here, $M_s$ stands for the saturation
magnetization. 
The full micromagnetic energy of the nanotube is given 
by \cite{Aharoni}
\begin{align}
  E(\vm) = &A \int_\Omega |\nabla \vm|^2 \ud \vx + K \int_\Omega
  \left[ 1 - (\vm \cdot \ex)^2 \right] \ud \vx
  \nonumber\\ &+\frac{\mu_0 M_s^2}{2} \int_{\mathbb{R}^3} |\nabla u|^2
  \ud \vx \,,
\label{eq:01}
\end{align}
where the magnetostatic potential $u(\vx,t)$ satisfies
\begin{equation}
  \nabla \cdot (\nabla u + \vm) = 0 \quad \mathrm{for} \quad \vx \in
  \mathbb{R}^3 \,.
\label{eq:02}
\end{equation}
%
Here, $A$ denotes the exchange constant, $K$ is the easy axis
anisotropy constant, $\mu_0 = 4 \pi \times 10^{-7}$ Wb$/($A$\cdot$m) is the
magnetic permeability of vacuum, and $\ex$ is a unit vector pointing
along the symmetry axis ($x$-axis) of the nanotube (see
Eq.~\ref{fig1}).

Within a continuum description, the time evolution of the
magnetization distribution is governed by the Landau-Lifshitz (LL)
equation \cite{LandauLifshitz35,Gilbert55}
\begin{equation}
  \frac{\partial \vm}{\partial t} = \gamma \vm \times \vH - \alpha \vm
  \times (\vm \times \vH) \,.
\label{eq:04}
\end{equation}
Here, $\gamma$ denotes the  
{
gyromagnetic ratio, 
} $\alpha$ is a
phenomenological damping parameter, and $\vH$ is an effective magnetic
field, given by
\begin{equation}
  \vH(\vm) = - \frac{1}{\mu_0 M_s} \frac{\delta E}{\delta \vm} + \vH_a \,,
\label{eq:03}
\end{equation}
where $\vH_a$ stands for the applied (external) magnetic field. Being
interested in the dynamics of a magnetization domain wall (DW), we
focus on solutions of Eq.~(\ref{eq:04}) subject to the boundary
conditions $\vm(\vx,t) \rightarrow \pm \ex$ for $x \rightarrow \pm
\infty$ (and $\vx \in \Omega$).

We now address the case of a thin nanotube, such that $w \ll R$. In
this limit, the volume integrals in Eq.~(\ref{eq:01}) can be
approximately reduced to integrals over the surface of a cylinder, 
{
and the stray-field energy can be approximated by
an additional effective local anisotropy that penalises the magnetisation component in the radial direction  (see
\cite{Carbou, KS05Thin} for mathematical details of this procedure).} Thus,
rescaling the spatial variables,  ${\bf x} = R \boldsymbol{\xi}$; the
micromagnetic energy,  $E = 2 A w \mathcal{E}$; and the effective and
applied fields, { $\vH = [2 A  /( \mu_0 M_s R^2)] \vcH$ and $\vH_a = [2 A / (\mu_0 M_s R^2)]
\vcH_a$, }we approximate Eqs.~(\ref{eq:01}--\ref{eq:03}) by
\begin{align}
  \mathcal{E}(\vm) = &\frac{1}{2} \int_S |\nabla_S \vm|^2 \ud \sigma +
  \frac{\kappa}{2} \int_S \left[ 1 - (\vm \cdot \ex)^2 \right] \ud
  \sigma \nonumber\\ &+ \frac{\lambda}{2} \int_S (\vm \cdot \erho)^2
  \ud \sigma
\label{eq:05}
\end{align}
and
\begin{align}
  \vcH(\vm) = \nabla_S^2 \vm + \kappa (\vm \cdot \ex) \ex - \lambda
  (\vm \cdot \erho) \erho + \vcH_a \,,
\label{eq:06}
\end{align}
where $\kappa = K R^2 / A$ and $\lambda = \mu_0 M_s^2 R^2 / (2
A)$. 
The integrals in Eq.~(\ref{eq:05}) run over the surface of an
infinitely long cylinder of  unit radius, and $\nabla_S = \ex
\frac{\partial}{\partial \xi} + \epsi \frac{\partial}{\partial \psi}$
represents the surface gradient (and, accordingly, $\nabla_S^2 =
\frac{\partial^2}{\partial \xi^2} + \frac{\partial^2}{\partial
  \psi^2}$ the surface Laplacian). Consequently, rescaling the time
variable as { $t = [ \mu_0 M_s R^2/(2 \gamma A ) ] \tau$}, we rewrite the LL
equation (\ref{eq:04}) in the dimensionless form,
\begin{equation}
  \frac{\partial \vm}{\partial \tau} = \vm \times \vcH -
  \frac{\alpha}{\gamma} \vm \times (\vm \times \vcH) \,.
\label{eq:07}
\end{equation}

Equations~(\ref{eq:05}--\ref{eq:07}), along with the boundary
condition $\vm(\xi, \tau) \rightarrow \pm \ex$ as $\xi \rightarrow \pm
\infty$ 
{
specify
the magnetization dynamics problem addressed in this paper. 
}
Below, we
provide exact, traveling wave solutions to this problem in the two
limiting cases of $\lambda \gg 1$ and $\lambda \ll 1$.

\medskip

\begin{figure}[ht]
\includegraphics[width=3in]{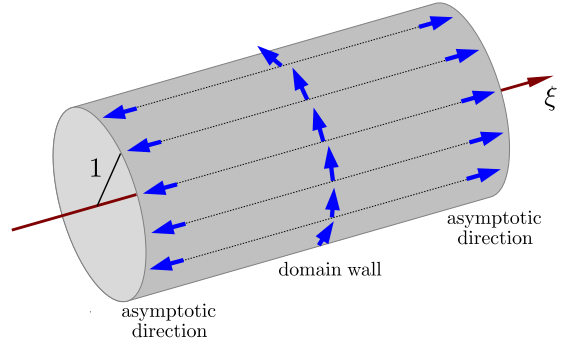}
\caption{(Color online) A sketch of a magnetization DW for the case of
  $\lambda \gg 1$.}
\label{fig2}
\end{figure}

{\it Strong penalization case, $\lambda \gg 1$. --} In ferromagnetic
nanotubes with very large $\lambda$, the penalization term in the
micromagnetic energy, Eq.~(\ref{eq:05}), essentially forces the
magnetization distribution $\vm$ to lie 
{
 nearly tangent to the}
cylinder (see Fig.~\ref{fig2}). More specifically, it can be shown
that 
$
\vm = \vm_{\mathrm{t}} + \lambda^{-1} \vm_{\mathrm{n}}$, where
$\vm_{\mathrm{t}} = (\vm \cdot \ex) \ex + (\vm \cdot \epsi) \epsi$
{
is tangent to the cylinder}
and $ \vm_{\mathrm{n}} = (\vm\cdot \erho) 
\erho$ (with $|\vm_{\mathrm{n}}| \sim \mathcal{O}(1)$) is normal to
the cylinder surface. 
%
%
%

Resolving the effective field into its tangential and normal
components 
$\vcH_{\mathrm{t}} = (\vcH \cdot \ex) \ex + (\vcH \cdot \epsi) \epsi$
and $\vcH_{\mathrm{n}} = (\vcH \cdot \erho) \erho$ (both
$|\vcH_{\mathrm{t}}|$ and $|\vcH_{\mathrm{n}}|$ being of order 1), we
rewrite Eq.~(\ref{eq:07}) as $\frac{\ud}{\ud \tau} \vm_{\mathrm{t}} =
\vm_{\mathrm{t}} \times (\vcH_{\mathrm{t}} + \vcH_{\mathrm{n}}) -
\frac{\alpha}{\gamma} \vm_{\mathrm{t}} \times [ \vm_{\mathrm{t}}
  \times (\vcH_{\mathrm{t}} + \vcH_{\mathrm{n}})) ] +
\mathcal{O}(\lambda^{-1})$. Then, 
{
resolving this equation into its tangential and normal components
}
and keeping terms of the leading order in
 {
 $\lambda^{-1}$,
 }
 we obtain
\begin{align}
  \frac{\ud}{\ud \tau} \vm_{\mathrm{t}} &= \vm_{\mathrm{t}} \times
  \vcH_{\mathrm{n}} - \frac{\alpha}{\gamma} \vm_{\mathrm{t}} \times (
  \vm_{\mathrm{t}} \times \vcH_{\mathrm{t}} ) \,, \label{eq:08a}\\ 0
  &= \vm_{\mathrm{t}} \times \vcH_{\mathrm{t}} - \frac{\alpha}{\gamma}
  \vm_{\mathrm{t}} \times ( \vm_{\mathrm{t}} \times
  \vcH_{\mathrm{n}} \label{eq:08b} ) \,.
\end{align}
Taking the cross product of both sides of Eq.~(\ref{eq:08b}) with
$\vm_{\mathrm{t}}$, and using $|\vm_{\mathrm{t}}|^2 = 1 +
 \mathrm{O}(\lambda^{-2})$
we obtain, to the leading order in
 {
 $\lambda^{-1}$,
 }
\begin{equation}
  \vm_t \times \vcH_{\mathrm{n}} = -\frac{\gamma}{\alpha}
  \vm_{\mathrm{t}} \times (\vm_{\mathrm{t}} \times \vcH_{\mathrm{t}})
  \,.
\label{eq:09}
\end{equation}
Finally, substituting Eq.~(\ref{eq:09}) into Eq.~(\ref{eq:08a}), we
conclude that, in the limit 
 {
$\lambda \rightarrow \infty$ 
 }
(or, more
generally, in the leading order in
{$\lambda^{-1}$) 
}
the time evolution of
$\vm(\xi,\psi,\tau)$ is governed by the modified LL equation,
\begin{equation}
  \frac{\partial \vm}{\partial \tau} = -\left( \frac{\alpha}{\gamma} +
  \frac{\gamma}{\alpha} \right) \vm \times (\vm \times
  \vcH_{\mathrm{t}}) \,,
\label{eq:10}
\end{equation}
where the magnetization distribution is 
{
restricted to be tangent to the surface of the cylinder,
}
\begin{equation}
  \vm = \ex \cos \theta + \epsi \sin \theta \,.
\label{eq:11}
\end{equation}
In general, $\theta = \theta(\xi,\psi,\tau)$. A similar result has
been obtained for the effective dynamics in thin ferromagnetic films
\cite{KS05Effective}.

We now assume that the applied magnetic field is directed along the
nanotube axis, $\vcH_a = \mathcal{H}_a \ex$. Substituting
Eq.~(\ref{eq:11}) into Eq.~(\ref{eq:06}), taking into account the fact
that $\frac{\partial}{\partial \psi} \epsi = -\erho$ and
$\frac{\partial}{\partial \psi} \erho = \epsi$, and discarding the
component of $\vcH$ along $\erho$, we obtain the tangential component
of the effective field,
\begin{align}
  \vcH_{\mathrm{t}} = &\big( -\sin \theta \, \nabla_S^2 \theta - \cos
  \theta \, |\nabla_S \theta|^2 + \kappa \cos \theta + \mathcal{H}_a
  \big) \ex \nonumber\\ &+ \big( \cos \theta \, \nabla_S^2 \theta -
  \sin \theta \, |\nabla_S \theta|^2 - \sin \theta \big) \epsi
  \,. \label{eq:12}
\end{align}
Consequently,
\begin{align}
  &\vm \times (\vm \times \vcH_{\mathrm{t}}) = \nonumber\\ &\big(
  \nabla_S^2 \theta - (1+\kappa) \sin \theta \cos \theta -
  \mathcal{H}_a \sin \theta \big) (\ex \sin \theta - \epsi \cos
  \theta) \,. \label{eq:13}
\end{align}
Thus, using the identity $\frac{\partial}{\partial \tau} \vm = -(\ex
\sin \theta - \epsi \cos \theta) \frac{\partial}{\partial \tau}
\theta$ and Eq.~(\ref{eq:13}) in the left- and right-hand side of
Eq.~(\ref{eq:10}) respectively, we obtain
\begin{equation}
  \frac{\partial \theta}{\partial \tau} = \left( \frac{\alpha}{\gamma}
  + \frac{\gamma}{\alpha} \right) \big( \nabla_S^2 \theta - (1+\kappa)
  \sin \theta \cos \theta - \mathcal{H}_a \sin \theta \big) \,.
\label{eq:14}
\end{equation}

Equation~(\ref{eq:14}) governs the dynamics of the magnetization
distribution, given by Eq.~(\ref{eq:11}), subject to the boundary
conditions $\lim\limits_{\xi \rightarrow -\infty} \theta
(\xi,\psi,\tau) = \pi$ and $\lim\limits_{\xi \rightarrow +\infty}
\theta (\xi,\psi,\tau) = 0$. It can be straightforwardly verified that
this problem admits a family of exact traveling wave solutions
\begin{equation}
  \theta(\xi,\psi,\tau) = \Theta_1 \big(\xi - \xi_0(\tau)\big) \,,
\label{eq:14.5}
\end{equation}
where the function
\begin{equation}
  \Theta_1 (\xi) = 2 \tan^{-1} \exp \left(- \xi \sqrt{1 + \kappa}
  \right)
\label{eq:15}
\end{equation}
(or, equivalently, $\frac{\ud}{\ud \xi} \Theta_1 = -\sqrt{1 + \kappa}
\, \sin \Theta_1$) determines the spatial profile of the traveling
wave, and
\begin{equation}
  \frac{\ud \xi_0}{\ud \tau} = -\left( \frac{\alpha}{\gamma} +
  \frac{\gamma}{\alpha} \right) \frac{\mathcal{H}_a}{\sqrt{1 +
      \kappa}}
\label{eq:16}
\end{equation}
gives the propagation velocity. In the original physical coordinates,
the propagation velocity reads
\begin{equation}
  \frac{\ud x_0}{\ud t} = -\left( \frac{\alpha}{\gamma} +
  \frac{\gamma}{\alpha} \right) \frac{\gamma R H_a}{\sqrt{1
      + K R^2 / A}} \,.
\label{eq:17}
\end{equation} 
Equation~(\ref{eq:17}) gives explicitly the nonlinear dependence of
the DW propagation speed on the nanotube radius. {Thus, in the
anisotropic case ($K > 0$), our formula shows that $|\frac{\ud}{\ud t}
x_0| \propto R H_a$ for $R \ll \sqrt{A/K}$, and $|\frac{\ud}{\ud t} x_0|
\propto H_a$ for $R \gg \sqrt{A/K}$. In the isotropic case ($K = 0$),
however, $|\frac{\ud}{\ud t} x_0 | \propto R H_a$ at any nanotube radius.}

\medskip

\begin{figure}[ht]
\includegraphics[width=3in]{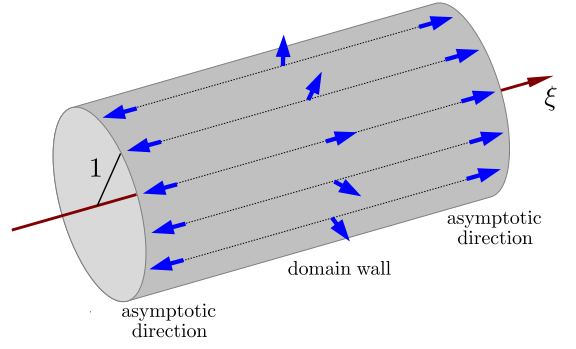}
\caption{(Color online) A sketch of the magnetization DW for the case
  of $\lambda \ll 1$. The DW has the helicity index $n=1$.}
\label{fig3}
\end{figure}

{\it Weak penalization case, $\lambda \ll 1$. --} We now focus on the
case of a ferromagnetic nanotube for which the penalization parameter
$\lambda$ is negligibly small. In this case the magnetization
distribution $\vm$ is no longer restricted to 
{
lie tangent to the cylinder}
and explores the full unit sphere. Its time evolution is
governed by the LL equation (\ref{eq:07}) with the effective field
approximated by (cf. Eq.~(\ref{eq:06}))
\begin{align}
  \vcH = \nabla_S^2 \vm + (\kappa \vm \cdot \ex + \mathcal{H}_a) \ex
  \,.
\label{eq:18}
\end{align}

Substituting the Cartesian representation of the magnetization
distribution, $\vm = (\cos \theta, \, \sin \theta \cos \phi, \, \sin
\theta \sin \phi)$, into Eqs.~(\ref{eq:07}) and (\ref{eq:18}), we
obtain a system of two coupled nonlinear PDEs for the unknown
functions $\theta = \theta(\xi,\psi,\tau)$ and $\phi =
\phi(\xi,\psi,\tau)$:
\begin{align}
  \frac{\partial \theta}{\partial \tau} + \frac{\gamma}{\alpha}
  \frac{\partial \phi}{\partial \tau} \sin \theta &= \left(
  \frac{\alpha}{\gamma} + \frac{\gamma}{\alpha} \right) \mathcal{F}_1
  \,, \label{eq:19a}\\ -\frac{\gamma}{\alpha} \frac{\partial
    \theta}{\partial \tau} + \frac{\partial \phi}{\partial \tau} \sin
  \theta &= \left( \frac{\alpha}{\gamma} + \frac{\gamma}{\alpha}
  \right) \mathcal{F}_2 \,, \label{eq:19b}
\end{align}
where
\begin{align}
   \mathcal{F}_1 = &\frac{\partial^2 \theta}{\partial \xi^2} +
   \frac{\partial^2 \theta}{\partial \psi^2} - \left[\kappa + \left(
     \frac{\partial \phi}{\partial \xi} \right)^2 + \left(
     \frac{\partial \phi}{\partial \psi} \right)^2 \right] \sin\theta
   \cos\theta \nonumber\\ &-\mathcal{H}_a \sin\theta
   \,, \label{eq:20a}\\ \mathcal{F}_2 = &2 \left[ \frac{\partial
       \theta}{\partial \xi} \frac{\partial \phi}{\partial \xi} +
     \frac{\partial \theta}{\partial \psi} \frac{\partial
       \phi}{\partial \psi} \right] \cos\theta + \left[
     \frac{\partial^2 \phi}{\partial \xi^2} + \frac{\partial^2
       \phi}{\partial \psi^2} \right] \sin\theta \,. \label{eq:20b}
\end{align}
{
As before, this system is to be solved subject to the boundary
conditions $\lim\limits_{\xi \rightarrow -\infty} \theta
(\xi,\psi,\tau) = \pi$ and $\lim\limits_{\xi \rightarrow +\infty}
\theta (\xi,\psi,\tau) = 0$ .}

As can be readily verified by a direct substitution, this problem
admits a two-parameter family of exact traveling wave solutions
\begin{align}
  &\theta(\xi,\psi,\tau) = \Theta_n \big( \xi - \xi_0(\tau) \big)
  \,, \label{eq:21a}\\ &\phi(\xi,\psi,\tau) = n \psi + \Phi(\tau)
  \,, \label{eq:21b}
\end{align}
with $n \in \mathbb{Z}$. Here, the longitudinal profile of the DW is
given by
\begin{equation}
  \Theta_n (\xi) = 2 \tan^{-1} \exp \left(- \xi \sqrt{n^2 + \kappa}
  \right)
\label{eq:22}
\end{equation}
(or, equivalently, $\frac{\ud}{\ud \xi} \Theta_n = -\sqrt{n^2+\kappa}
\, \sin \Theta_n$), the precession velocity by
\begin{equation}
  \frac{\ud \Phi}{\ud \tau} = -\mathcal{H}_a \,,
\label{eq:23}
\end{equation}
and the propagation velocity by
\begin{equation}
  \frac{\ud \xi_0}{\ud \tau} = -\frac{\alpha}{\gamma}
  \frac{\mathcal{H}_a}{\sqrt{n^2+\kappa}} \,.
\label{eq:24}
\end{equation}
In the original physical coordinates, the propagation velocity reads
\begin{equation}
  \frac{\ud x_0}{\ud t} = -\frac{\alpha 
    R H_a}{\sqrt{n^2 + K R^2 / A}} \,.
\label{eq:25}
\end{equation}
In Eqs.~(\ref{eq:21a}--\ref{eq:25}), the index $n$ measures the DW
helicity. That is, $n$ counts the number of times that the
magnetization vector 
{
turns about $\ex$ as  the circumference of the cylinder is traversed.}
(A sketch of a DW with $n=1$ is shown in Fig.~\ref{fig3}.) It
is interesting to note that DWs with lower helicity propagate faster,
with the maximal propagation speed, $|\frac{\ud}{\ud \tau} \xi_0| =
(\alpha / \gamma) |\mathrm{H}_a| / \sqrt{\kappa}$, achieved for $n =
0$.

As in the strong penalization case, Eq.~(\ref{eq:25}) gives the full
nonlinear dependence of the DW propagation speed on the nanotube
radius. {In the anisotropic case ($K > 0$), we see that
$|\frac{\ud}{\ud t} x_0| \propto R H_a$ for $R \ll n \sqrt{A/K}$, while
$|\frac{\ud}{\ud t} x_0| \propto H_a$ for $R \gg n \sqrt{A/K}$. In the
isotropic case ($K = 0$), we again recover the scaling $|\frac{\ud}{\ud
  t} x_0| \propto R H_a$.}

\medskip

In conclusion, we have conducted an analytic study of the DW dynamics
in thin ferromagnetic nanotubes 
{ subject to} 
external longitudinal
magnetic fields. We have found explicit functional forms of the DW
profiles and derived explicit formulas for the DW velocity in the
regimes of strong and weak penalization, Eqs.~(\ref{eq:17}) and
(\ref{eq:25}) respectively. In the strong penalization case, the
magnetization field lies { nearly tangent to the}
nanotube, while for weak penalizations, the magnetization vector may
wrap around the nanotube with any integer helicity index. The DW
propagation speed increases with the nanotube radius in a nonlinear
way, and, in the weak penalization case, decreases with increasing
helicity. Since for a typical ferromagnetic material $\alpha/\gamma
\ll 1$, DWs in the strong-penalization case propagate much faster than
those in the weak-penalization case.

\emph{Acknowledgments.--} A.G.~thanks EPSRC for support under grant  EP/K024116/1, J.M.R.~thanks EPSRC for support under grant  EP/K02390X/1, and V.S.~thanks EPSRC for support under grants  EP/I028714/1 and  EP/K02390X/1.

\end{document}